\documentclass[reprint,aps,prl,groupedaddress,floatfix]{revtex4-2}
\usepackage{graphicx}
\usepackage{amsmath}
\usepackage{amssymb}
\usepackage{braket}
\usepackage{eqnarray}
\usepackage{dsfont}
\usepackage{multirow}
\usepackage{circledsteps}
\usepackage[T1]{fontenc}

\newcommand\sqnorm{\frac{1}{\sqrt{2}}}

\newcommand\plink{p_\text{link}}
\newcommand\ploop{p_\text{loop}}

\begin{document}

\title{Long-distance quantum communication sending
single photons and keeping many}
\author{Stefan H{\"a}ussler and Peter van Loock}
\affiliation{Institute of Physics, Johannes Gutenberg-Universit{\"a}t Mainz, Staudingerweg 7, 55128 Mainz, Germany}
\date{\today}

\begin{abstract}
Fiber-based classical communication is all-optical and uses light pulses reamplfied and reshaped
every 50-100 km in classical repeaters. Most compatible with this would be a quantum 
communication system which is also all-optical with quantum processing units placed in similar intervals.
However, existing all-optical quantum communication protocols either require complicated quantum error correction steps for logical-qubit recoveries at every few kilometers  or, over larger quantum repeater segments, they would at least depend on sharing complex multi-photon entangled states.
Here we propose an all-optical memory-based quantum repeater for long-distance quantum communication, with quantum memories at each repeater station realized in the form of fiber loops combined
with suitable quantum error correction codes for photon-loss protection.
By sending only single-photon states through the fibers connecting the stations, such repeaters can operate in the classical infrastructure's long-segment regime. We analyze the performance of our scheme for the 
Gottesman-Kitaev-Preskill code, including a concatenation with the Steane code, as well as the single-photon quantum parity
code for total distances up to 10000km.
\end{abstract}

\maketitle

\textit{Introduction. }
To enable quantum key distribution (QKD) \cite{BB84} over practically relevant distances, the fundamental limit \cite{plob} imposed on transmission rates by the exponentially decreasing quantum capacity of the bosonic loss channel must be overcome. A promising approach is the idea of quantum repeaters (QR), that is to introduce intermediate stations along the transmission line, and thus partitioning the total distance $L$ into $n$ shorter segments of length $L_0 = L/n$. In the time since the first repeater proposal \cite{firstrepeater}, a multitude of other schemes have been put forward \cite{secondrepeater, Azuma2015, Muralidharan2014, Ewert2016, Ewert2017, Lee2019, alloptical, Rozpedek2021, Rozpedek2023, thirdgen_surfacecode, thirdgen_gkp, häussler_vanloock}. They can be categorized into so-called repeater generations \cite{generations}, or more broadly into memory-based and memoryless protocols. While the latter can theoretically achieve higher rates under favorable conditions due to their repetition rate not being restricted by classical communication times, they are also more sensitive to imperfections, and in particular require a very high fiber-coupling efficiency as well as a rather close station spacing ($\sim$ 1--10 km). Contrarily, the memory-based approach enables a non-vanishing transmission rate even at spacings in the order of 100 km and medium link efficiencies.
In principle, any stationary quantum system can act as a quantum memory; however, matter-based memories with sufficient coherence times, be it in the form of atomic spin ensembles, superconducting circuits or ions trapped in electromagnetic fields, are very challenging to implement and control in practice.

In this work we propose a QR using optical storage in form of fiber loops of length $L_\text{loop} = L_0/m$, for some $m\in\mathds{N}$, located in the intermediate stations, together with a quantum error correction (QEC) code, such that encoded logical states exiting the loop are teleported into fresh states entering the loop again.  
Making use of QEC on the stored states is crucial for our approach, as for single-photon states, the achievable transmission rates would lie below the repeaterless bound \cite{plob}: Even in the improbable case of the entanglement distribution attempts succeeding on the first try in all segments, states in two memories at each of the $n-1$ stations would have to undergo at least $m$ teleportations while waiting for the signal heralding distribution success, which, due to the teleportation success probability scaling with the loop's transmissivity $\eta_\text{loop} \propto \exp(-L_\text{loop}/L_\text{att})$ where $L_\text{att} = 22$km denotes the fiber's attenuation length, results in the rate being upper-bounded by $\exp(-L/L_\text{att})^{2m(n-1)/nm}$. Already with $n > 2$, this lies below the transmissivity $\eta = \exp(-L/L_\text{att})$ of a fiber spanning the total communication distance, and thus also below the PLOB-bound \cite{plob} of $-\log_2(1 - \eta)$. 

To overcome this issue, we consider in this work two QEC codes of very different flavors: On the one hand, the bosonic Gottesman-Kitaev-Preskill (GKP) code \cite{gkp} that encodes a qubit into a grid-like structure in a bosonic mode's phase space in order to offer protection against random displacement errors, and on the other hand the ``fermionic'' quantum parity code (QPC) \cite{qpc}, where a qubit is encoded into multiple modes, each containing either one or zero photons.
The Bell-state measurements (BSM) used to teleport exiting states back into the loop simultaneously enable repeated teleportation-based error correction during the storage time, thus increasing the effective coherence time experienced by the logical states over the fixed value of about $1.1\times 10^{-4}$s governed by the fiber attenuation.
While the idea of fiber loops used as quantum memories is not a new one \cite{Pittman2002, Pittman2002b, Saglamyurek2015, Kim2024, Rozpedek2023}, its application in the context of a so-called second-generation QR, where memories serve the purpose of saving successfully distributed entanglement while waiting for distribution success in adjoining segments, in combination with the aforementioned QEC codes, constitutes a new contribution.
Other forms of all-optical quantum memories, such as resonators \cite{Yoshikawa2013, Hanamura2025} and Herriott cells \cite{Herriott1964, Herriott1965, Robert2007, Ou2023, Guo2025}, do not currently offer sufficient storage times and transmission efficiencies; however, such concepts have the potential to facilitate all-optical memory-based quantum communication with easier QEC codes if their performance improves in the future.  
We would like to stress that the concept
of an all-optical QR as proposed in this work is independent of  
a specific QEC code, like the two distinct examples GKP and QPC  
analyzed in our work. As elementary optical resources,
all what is needed are ``hybrid'' entangled pairs of a physical (dual-rail)
single-photon qubit and a logical multi-photon qubit \cite{hybrid_entanglement, Chen2002, Park2012, Kwon2013, Morin2014, Darras2023, Cavailles2018, Andersen2015, vanloock2011}
in a protocol where the single photons are sent through the optical fiber
channels connecting the repeater stations and where the multi-photon
states are stored at the stations.
We believe this proposal to be conceptually attractive, 
because its kind of second-generation memory-based two-way architecture 
allows for a much larger station spacing than
existing all-optical one-way (third-generation) QR schemes \cite{Ewert2016, Ewert2017, Lee2019}, while sending only single photon states
through the fiber and having the more complex encoded
states restricted to the local memories, where they can
be better manipulated and controlled.
This approach is thus also conceptually different from
all-photonic two-way QR schemes in which multi-photon
entangled cluster states are distributed between the stations \cite{Azuma2015}.

\begin{figure}
\includegraphics[width=0.45\textwidth]{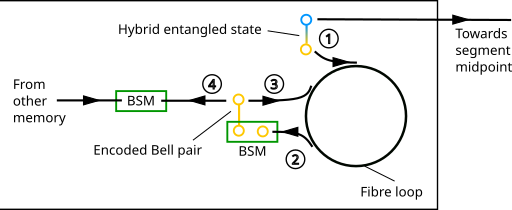}
\caption{(Color online) Sketch of the all-optical memory: \Circled{1} The encoded (yellow) half of the hybrid entangled state of Eq.~(\ref{eq::hybridstate}) is coupled into the fiber loop for storage, while the dual-rail qubit travels through the fiber towards the segment midpoint. \Circled{2} During the wait time, intermediate corrections are applied via a Bell-state measurement (BSM) together with a fresh encoded Bell pair, \Circled{3} after which the Bell pair's other component is coupled back into the loop. \Circled{4} Upon receiving the signal heralding distribution success in the adjoining segment to the left, the state is directed towards a second BSM device, where it is combined with the state stored in the station's other memory for entanglement swapping. }
\label{fig::setup}
\end{figure}
\textit{All-optical quantum memory. }
In the following, we go over the working principle of all-optical storage of encoded quantum states in a fiber loop, describing the memory's setup and its application in a quantum repeater context.
The basis for a second-generation QR is formed by a hybrid entangled state of the form 
\begin{equation}
\label{eq::hybridstate}
\sqnorm\left(\ket{\overline{0}}\ket{H} + \ket{\overline{1}}\ket{V}\right),
\end{equation}
where the first half is a logical qubit encoded in an arbitrary QEC code (indicated by the ``bar'' above the state designation), and the second half corresponds to a polarization-encoded photonic dual-rail qubit with $H$ denoting horizontal and $V$ denoting vertical polarization. After state generation, the single-photon component is coupled into the fiber and sent towards the segment's midpoint, where a BSM is performed together with the photon arriving from the segment's other endpoint in order to establish entanglement between the two respective encoded components.  
This entanglement distribution process succeeds with a probability of 
$
p = p^\text{(BSM)}\plink^2 e^{-\frac{L_0}{L_\text{att}}},
$
where $\plink$ is related to the fiber coupling efficiency and $p^\text{(BSM)}$ denotes the BSM success probability conditioned on both photons being present at the inputs which for linear-optics implementations is typically limited by $1/2$,
and the number of attempts until success forms a geometric distribution.
Meanwhile, as illustrated in Fig.~\ref{fig::setup}, the encoded component is coupled into the fiber loop for storage at least until the classical signal from the midpoint arrives carrying the information about whether the BSM was successful or not. In case of failure, the state is switched out of the loop and discarded, whereas in case of success it is kept in storage, since it might have to wait for the neighboring segment to also finish distribution before entanglement swapping can take place.  
During storage, teleportation-based error correction is applied to the encoded state after each pass through the loop by means of a logical BSM together with a freshly generated logical Bell pair, in order to counteract the effect of fiber attenuation, modeled as a bosonic loss channel with transmissivity
$
\eta_\text{loop} = \ploop e^{-\frac{L_0}{mL_\text{att}}},
$
where $\ploop$ similarly to $\plink$ accounts for the imperfect efficiency of coupling into and out of the fiber, but now for the encoded state instead of for a single photon, and $m = L_0/L_\text{loop}$ denotes the number of loop iterations during one classical signaling period.

Depending on the specific QEC code, each teleportation either introduces a logical Pauli error with a certain small, but non-vanishing probability (e.g. for the GKP code), or reduces the overall transmission rate due to the BSM not being deterministic (e.g. for the QPC).
In a repeater chain with $n$ segments, $n-1$ entanglement swappings will take place; however, the number of intermediate correction steps is given by the random variable 
$
\mathcal{M}_n = mD_n + 2m(n-1),
$
where $D_n$ is the random variable corresponding to the number of timesteps for which segments are waiting for other segments summed over the entire chain (similar to Ref.~\cite{rateanalysis}, but under the assumption of immediate measurements at Alice and Bob and differing by a factor of $2$. For a more detailed discussion of the distinction, see appendix of Ref.~\cite{häussler_vanloock}).

Typically, a repeater's performance is quantified in terms of the secret key rate (SKR) $S=rR$, composed of the raw rate $R$ and the secret key fraction (SKF) $r$.
However, in this work, we restrict our attention to the latter, since the raw rate \cite{Bernardes2011, Shchukin2019}
\begin{equation}
R = \left[\tau_0\sum_{i=1}^n (-1)^{i+1}\begin{pmatrix}n\\i\end{pmatrix}\frac{1}{1-q^i }\right]^{-1},
\end{equation}
with $q = 1-p$ and $\tau_0 = L_0/c$
is determined solely by the external parameters, i.e. number and length of segments, as well as the link efficiency $\plink$, and does not depend on the choice of QEC code used on the memories. Note that in our convention, the raw rate captures only the effect of the probabilistic entanglement distribution, while other non-unit probabilities, e.g. from entanglement swapping in QPC repeaters, are absorbed into the SKF.

\begin{figure}
\includegraphics[width=0.4\textwidth]{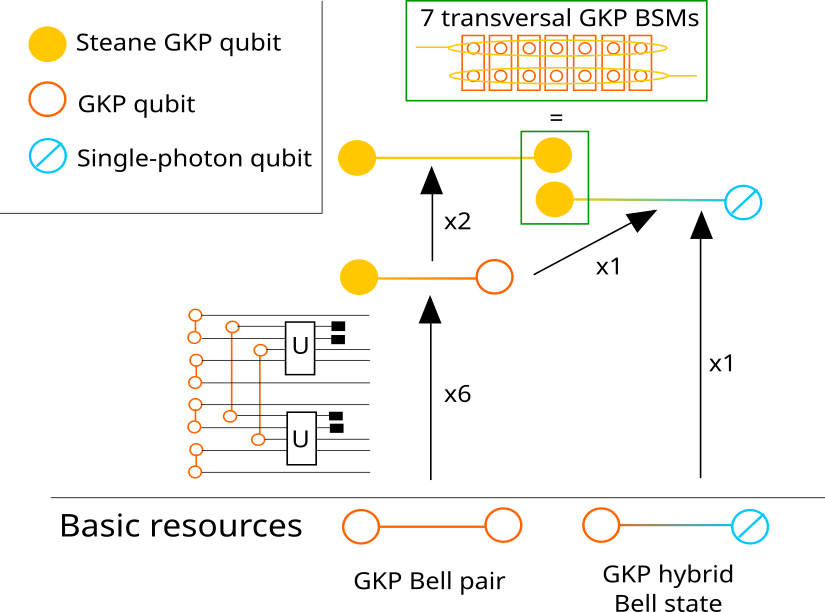}
\caption{(Color online) Use of basic resources for the Steane-GKP protocol: Six GKP Bell pairs \cite{Walshe2020} are converted by a passive linear-optics setup into a GKP Steane-GKP hybrid \cite{cubecreation}. States of this type can be used to create the entangled states used in the repeater scheme. Seven GKP BSMs provide the full Steane-GKP syndrome information \cite{Schmidt2022}.}
\label{fig::basicresources}
\end{figure}

\textit{GKP. }
\label{sec::GKP}
The GKP code \cite{gkp}, designed to protect logical information encoded in the Hilbert space of a bosonic mode from small random shifts in phase space, is defined as the subspace spanned by the idealized basis states
\begin{equation}
\label{eq::GKPdef}
\ket{\overline{0}} = \sum_{k \in \mathds{Z}}\Ket{2k\sqrt{\pi}}_q,\qquad
\ket{\overline{1}} = \sum_{k \in \mathds{Z}}\Ket{(2k + 1)\sqrt{\pi}}_q,
\end{equation}
where kets with index $q$ denote $q$-quadrature eigenstates. 
A feature making it an attractive candidate for all-optical storage is its suitability for deterministic logical BSMs implementable using beam splitters and homodyne detection. Each BSM, however, contributes a certain Pauli error probability depending on the loss experienced as well as a so-called squeezing parameter $s$ characterizing the initial state quality: Since the states as defined in Eq.~(\ref{eq::GKPdef}) are unphysical, and can only be approximately produced in practice, we model realistic GKP states via application of a Gaussian noise channel whose variance is inversely correlated to the squeezing parameter $s$ (see Supplemental Material for more details, as well as information about how encoded and hybrid Bell states can be generated for the QEC codes considered in this paper).
For the GKP code, the SKF of our repeater is given by
$
r = 1 - 2h(\epsilon),
$
with the binary entropy function $h(x) = -x\log_2x - (1-x)\log_2(1-x)$ and the quantum bit error rate (QBER) $\epsilon$.
The latter gives the probability of the final state containing an $X$ error (or equivalently a $Z$ error, since both occur with equal probability for the square-lattice GKP code) and is found by combining the elementary Pauli error probabilities $p^\text{Pauli}_\text{corr}$ and $p^\text{Pauli}_\text{swap}$ arising from any of the $\mathcal{M}_n$ intermediate corrections as well as from the $n-1$ swappings, respectively, via:
\begin{equation}
\label{eq::qber}
    \epsilon = \left[\mathds{E}\left(\left(\mathbf{p}^\text{Pauli}_\text{corr}\right)^{\circledast \mathcal{M}_n}\right) \circledast \left( \mathbf{p}^\text{Pauli}_\text{swap}\right)^{\circledast n-1}\right]_1.
\end{equation}
Therein, the symbol $\circledast$ is used to denote the circular convolution
\begin{equation}
    (\mathbf{a} \circledast\mathbf{b})_r = \sum_{i = 0}^{1} a_i b_{r - i \mod 2} ,
\end{equation}
of the zero-indexed lists $\mathbf{p}^\text{Pauli}_\text{corr} = [1 - p^\text{Pauli}_\text{corr}, p^\text{Pauli}_\text{corr}]$ and $\mathbf{p}^\text{Pauli}_\text{swap} = [1 - p^\text{Pauli}_\text{swap}, p^\text{Pauli}_\text{swap}]$.
The expectation value appearing in Eq.~(\ref{eq::qber}) as well as later Eq.~(\ref{eq::qpcskf}) can be approximately evaluated using the relation
\begin{equation}
\label{eq::expectation_approximation}
    \mathds{E}(a^{D_n}) \approx \left(\frac{1-q}{1+q}\frac{1+qa}{1-qa}\right)^{n-1}
\end{equation}
with the distribution failure probability $q = 1-p$.
For more details, see Ref.~\cite{häussler_vanloock} or the Supplemental Material.

\textit{Steane-GKP. }
In addition to the regular GKP code, we also consider a concatenation with the 7-qubit Steane code \cite{steane1, steane2}, where each of the individual ``physical'' qubits making up the outer Steane code is formed by a GKP qubit. The concatenation enables correction of some of the logical Pauli errors stemming from the GKP teleportations by additionally measuring the Steane-code stabilizers in each logical BSM. The SKF can be found similarly as before, with the difference that the elementary Pauli error probabilities must be passed through the transfer function
\begin{equation}
\label{eq::steanetransfer}
q_\text{Steane}^\text{Pauli} = (q_\text{GKP}^\text{Pauli})^7 + 7(q_\text{GKP}^\text{Pauli})^6(1 - q_\text{GKP}^\text{Pauli})
\end{equation}
before using them to determine the QBER via a slightly modified version of Eq.~(\ref{eq::qber}) that also accounts for Pauli errors arising from state generation. (see Supplemental Material). In Eq.~(\ref{eq::steanetransfer}), $q$ with sub- and superscripts refers to the complementary probability $1-p$ for $p$ carrying the same sub- and superscripts.
The Steane code is an attractive candidate for an outer code in conjunction with GKP qubits since both the state generation \cite{cubecreation} as well as the combined Bell-state and syndrome measurement \cite{Schmidt2022} can be implemented using linear optics.

\textit{QPC. }
A code with very different properties is the QPC \cite{qpc}, most easily expressed in the $X$-basis as
\begin{eqnarray}
    \ket{\overline{+}} &=& \frac{1}{\sqrt{2^n}}\left(\ket{0}^{\otimes a} + \ket{1}^{\otimes a}\right)^{\otimes b}\nonumber\\
    \ket{\overline{-}} &=& \frac{1}{\sqrt{2^n}}\left(\ket{0}^{\otimes a} - \ket{1}^{\otimes a}\right)^{\otimes b}
\end{eqnarray}
for any natural numbers $a$ and $b$.
Contrarily to the GKP code, no Pauli errors occur during teleportation; however, this comes at the cost of BSMs no longer being deterministic even in the no-loss case.
Thus, instead of the effect of a non-zero QBER, the SKF now captures the probability of all teleportations succeeding:
\begin{eqnarray}
\label{eq::qpcskf}
r &=& \left[1 - \left(\frac{1}{2}\right)^b\right]^{n-1}\mathds{E}\left(p_\text{QPC}^{\mathcal{M}_n}\right),
\end{eqnarray}
where the probability $p_\text{QPC}$ of successfully teleporting a lossy QPC qubit through a perfect QPC Bell pair is derived in Refs.~\cite{Ewert2016, Ewert2017} as 
\begin{equation}
\label{eq::qpcprob}
    p_\text{QPC} = \left[1 - (1 - \eta_\text{loop})^a\right]^b- \left[ 1 - (1 - \eta_\text{loop})^a - \frac{\eta_\text{loop}^a}{2}\right]^b.
\end{equation}

\textit{Results. }
\begin{figure}
    \includegraphics[width=0.4\textwidth]{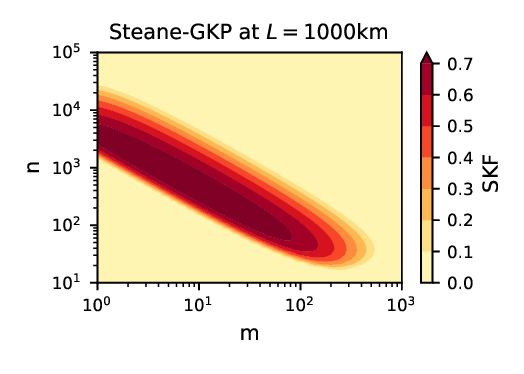}
    \caption{(Color online) SKF of Steane-GKP repeater as a function of loop and segment number, $m$ and $n$, respectively. Operating in the long-segment regime is possible with enough intermediate corrections. Parameters: $L=1000$km, $s=15$dB, $\plink=\ploop=0.99$. Pauli errors during state preparation are included in the SKF.}
    \label{fig::steane_skf}
\end{figure}
\begin{figure}
\includegraphics[width=0.4\textwidth]{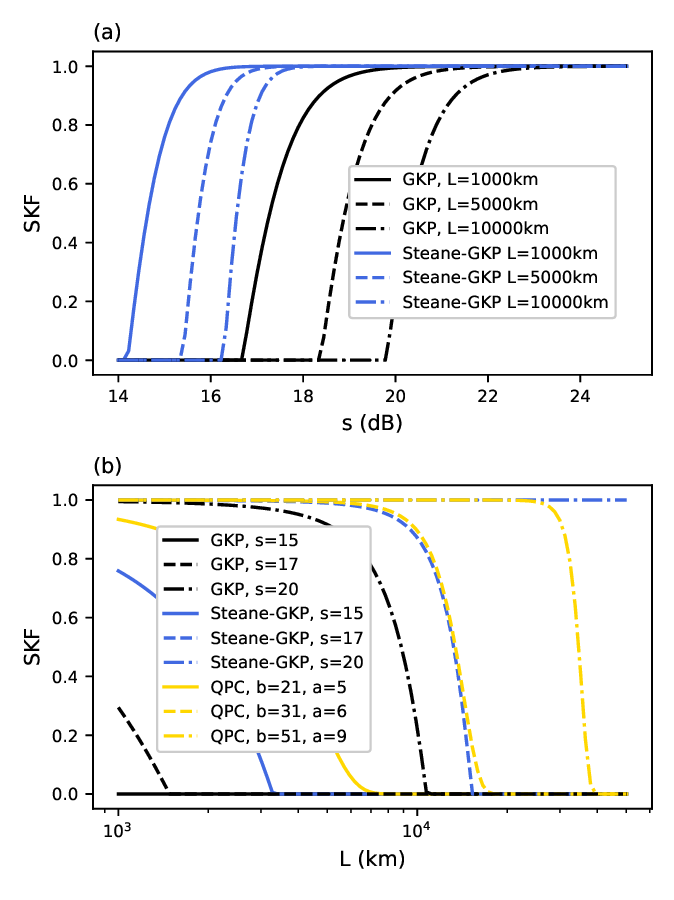}
\caption{(Color online) (a) Squeezing demands of regular GKP and Steane-GKP codes for different total lengths. (b) Secret key fractions as a function of total distance for various protocols. Parameters: $n=100$, $\plink=\ploop=0.99$, $m$ optimized. Plots for Steane-GKP include the effect of Pauli errors during state preparation.}
\label{fig::sLplot}
\end{figure}
When plotting the SKF over the $n$-$m$-plane for our repeater schemes, one observes that the area of non-zero SKF follows the shape of an oblong downward-sloping ellipse. This phenomenon is visualized in Fig.~\ref{fig::steane_skf} for a Steane-GKP repeater at a total distance of $L=1000$km and in the Supplemental Material for the other schemes, and gives rise to two distinct operational regimes near the extremal points: The first operates with few corrections steps and a large number of segments (upward of $10^3$), whereas the second is characterized by fewer but longer segments (up to $50$km in Fig.~\ref{fig::steane_skf}, but lengths $\geq$ $100$km are possible with higher squeezing) as well as more intermediate corrections. It is this second regime that is particularly interesting because, in contrast to the first, it is not directly accessible to schemes based on encoded entanglement distribution such as Ref.~\cite{Rozpedek2023}.
Thus, we believe our scheme to be the first repeater proposal combining the advantages of an all-optical architecture with those of being compatible with existing fiber-based communication infrastructure.

One of the major difficulties impeding the experimental realization of a GKP-based QR is posed by the challenge of creating sufficiently strongly squeezed GKP states. Finding ways to reduce the squeezing demand is therefore of vital importance for any practical implementation of such schemes.   
In Fig.~\ref{fig::sLplot}(a) we plot the achievable SKFs of the two GKP-based schemes for various external repeater configurations as a function of the squeezing parameter $s$ while optimizing over the number of intermediate correction steps $m$.
We find that with $100$ segments and fiber coupling efficiencies of $\plink=\ploop=0.99$, the regular GKP scheme requires squeezing in the range of $17$dB to $20$dB to cover distances between $1000$km and $10000$km, whereas concatenating with the Steane code reduces the required squeezing by $3$dB to $4$dB into the range of $14$dB to $16$dB.
In Fig.~\ref{fig::sLplot}(b) we include the QPC repeater in the comparison, now plotting SKF over the total distance. 
For the QPC, we choose the number of photons per block $a$ that we found to perform best for the given number of blocks $b$. 
With $100$ segments, almost all schemes under consideration, with the exception of the regular GKP at $15$dB squeezing, can achieve a non-zero SKF at $1000$km distance, most of them even a SKF of $1$. A distance of $10000$km can be reached by the Steane-GKP schemes with more than $15$dB squeezing and the QPC with $31$ or more blocks. The regular GKP with $20$dB squeezing also just about surpasses the $10000$km mark. The raw rate at $L=1000$km with $10$ segments is $3.5$Hz and at $L=10000$km with $100$ segments it is roughly $2$Hz; however, it may be possible to increase these values using multiplexing.

\textit{Conclusion. }
We have introduced an all-optical quantum repeater scheme using fiber-loop-based storage and correction of photonic logical qubits encoded in QEC codes.
Our scheme requires no complex multi-photon entangled states to be distributed in the fiber communication channels; instead, like in 
many existing experimental quantum communication demonstrations, only single photons need to be transmitted.
With a suitable number of intermediate correction steps and resource states of sufficient quality, our scheme can operate with long segments comparable to classical optical communication infrastructure.

\textit{Acknowledgments. }
We thank the BMFTR in Germany for
support via PhotonQ, QR.N, QuKuK, and QuaPhySI.
We further acknowledge support via the EU project
CLUSTEC (grant agreement no. 101080173).

\bibliography{literatur}
\bibliographystyle{apsrev4-1}

\clearpage
\appendix
\section*{Supplemental Material}

\subsection{GKP teleportation}
The basis states of the two-dimensional GKP code \cite{gkp} are given by
\begin{eqnarray}
\ket{\overline{0}} &=& \sum_{k \in \mathds{Z}}\Ket{2k\sqrt{\pi}}_q\nonumber\\
\ket{\overline{1}} &=& \sum_{k \in \mathds{Z}}\Ket{(2k + 1)\sqrt{\pi}}_q,
\end{eqnarray}
where kets with index $q$ denote $q$-quadrature eigenstates.
Being composed of displaced infinitely squeezed states, these states cannot be realized exactly in practice, but only be approximated by replacing delta-peaks in phase space with narrow Gaussians and truncating the infinite sum on both sides. In theoretical descriptions it is common to omit the truncation and instead modulate the sum by a wide Gaussian envelope; however, here we will make use of the simpler Gaussian noise approximation to account for finite squeezing effects: we define realistic GKP states as resulting from the action of a Gaussian displacement channel on the idealized states:
\begin{equation}
\left(\ket{\overline{j}}\bra{\overline{j}}\right)_\text{real} = \mathcal{E}_{\delta^2}\left[\left(\ket{\overline{j}}\bra{\overline{j}}\right)_\text{ideal}\right],
\end{equation}
with the Gaussian displacement channel with variance $\delta^2$ defined as
\begin{equation}
\label{eq::shiftchannel}
\mathcal{E}_{\delta^2}[\rho] = \frac{1}{\pi\delta^2}\int_\mathds{C} d^2\alpha\, \exp\left(-\frac{|\alpha|^2}{\delta^2}\right)D(\alpha)\rho D^\dagger(\alpha).
\end{equation}
Instead of referring directly to the variance, it is more common to use a squeezing parameter $s$ expressed in units of dB and defined as
\begin{equation}
    s = -10\log_{10}(2\delta^2).
\end{equation}

The GKP code is specifically designed to protect against random Gaussian displacements in phase space. However, since the dominant noise channel in our repeater is bosonic loss in the memories' fiber loops, an additional step is required to convert the loss into random Gaussian shifts. Such a conversion can be achieved using preamplification \cite{channels}, which turns the loss channel with transmissivity $\eta_\text{loop}$ into a shift channel with variance $\sigma^2_\text{loop} = 1 - \eta_\text{loop}$. Together with the finite-squeezing variance, this determines the probability of logical Pauli errors occurring during a teleportation step.
Note that the technique of CC (``classical computer'') amplification \cite{alloptical}, that can be performed without any physical amplification, is not applicable here since it requires losses of equal strength on the two modes on which the BSM is applied for teleportation. 
While such a symmetric-loss situation could be created by subjecting one half of the Bell pair to artificial loss matching the one experienced by the state traversing the loop, this is not advantageous: Both modes would contribute $\frac{1 - \eta_\text{loop}}{2\eta_\text{loop}}$ to the variance after CC-amplification \cite{alloptical}, resulting in $\sigma^2_\text{loop} = \frac{1 - \eta_\text{loop}}{\eta_\text{loop}}$, which is larger than the variance obtained with preamplification for all $\eta_\text{loop} \in [0 , 1]$.
Further note that not doing amplification at all is also a possible option that may even lead to an improved performance \cite{Hastrup2023}.

\begin{figure}
\includegraphics[width=\linewidth]{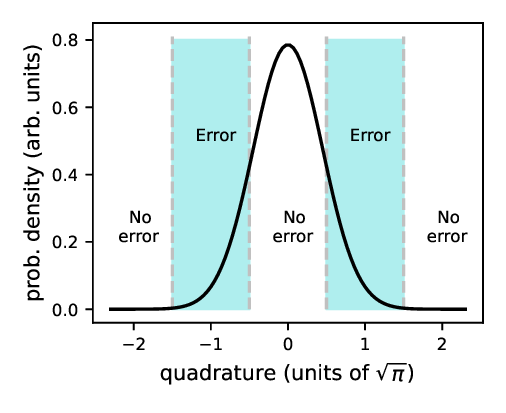}
\caption{(Color online) Error correction of Gaussian shifts. The graph shows the distribution of the true shift, the dashed lines separate shaded regions where the syndrome interpretation leads to a logical error from those where no error occurs.}
\label{fig::errorregions}
\end{figure}

The teleportation itself is carried out by mixing the data mode with one half of a GKP-encoded Bell pair at a beam splitter and subsequently measuring the output modes using homodyne detection. The measurement results, which we will denote $\overline{x}$ and $\overline{p}$, yield both the information necessary for teleportation, namely which Pauli operator to apply to the remaining mode in order to reproduce the original state, as well as the syndrome information for error correction.
In the following, we explain how the measurement result should be interpreted for the position quadrature. 
Since the Bell state's mode has not been subject to any loss channel, it will only contribute integer multiples of $\sqrt{\pi}$ to $\overline{x}$, and thus any possible deviations must have been caused by errors in the data mode (the finite-squeezing variance of the Bell mode can be pushed onto the data mode, such that the latter carries the variance $\sigma^2_\text{tot} = \sigma^2_\text{loop} + 2\delta^2$ while the former is assumed to be perfect). Hence the shift of the data mode can be estimated by choosing that representative of the equivalence class $[\overline{x} \mod \sqrt{\pi}]$ that lies between $-\frac{\sqrt{\pi}}{2}$ and  $\frac{\sqrt{\pi}}{2}$. Sufficiently small shifts can thus be detected correctly. 
If, however, the absolute value of the true shift is larger than  $\frac{\sqrt{\pi}}{2}$, this method of interpretation will yield an incorrect value, resulting in the corrected state differing from the original state by a shift of $k\sqrt{\pi}$, $k\in \mathds{Z}$. For odd $k$ this results in a logical Pauli-$X$ error, whereas for even $k$ theoretically no error occurs, as a shift by an even multiple of $\sqrt{\pi}$ maps the ideal code states onto themselves. This results in the striped pattern under the Gaussian with variance $\sigma^2_\text{tot} = \sigma^2_\text{loop} + 2\delta^2$ illustrated in Fig.~\ref{fig::errorregions}. 
A similar pattern emerges for the momentum quadrature and possible logical $Z$ errors. 
The Pauli error probability for intermediate correction steps is now given by the sum of integrals
\begin{equation}
\label{eq::paulierrorintegral}
p^\text{Pauli}_\text{corr} = \sum_{k\in\mathds{Z}} \int_{(2k+1)\sqrt{\pi}-\sqrt{\pi}/2}^{(2k+1)\sqrt{\pi} + \sqrt{\pi}/2}dx\, \frac{1}{\sqrt{2\pi\sigma_\text{tot}^2}}\exp\left(-\frac{x^2}{2\sigma_\text{tot}^2}\right),
\end{equation}
and with the definition of $\sigma^2_\text{tot}$ modified to simply $\sigma^2_\text{tot} =  2\delta^2$ this also gives the Pauli error probability $p^\text{Pauli}_\text{swap}$ for the entanglement swapping (where both modes are assumed to be lossless). 

The repeater's QBER is most elegantly expressed via circular convolution as in Eq.~(\ref{eq::qber}) of the main text; however, for qubits, it can also be written in a more explicit form:
\begin{equation}
\label{eq::qber_explicit}
\epsilon = p_\text{all corr}^\text{Pauli}(1 - p_\text{all swap}^\text{Pauli}) + p_\text{all swap}^\text{Pauli}(1 - p_\text{all corr}^\text{Pauli}),
\end{equation}
where $p_\text{all corr}^\text{Pauli}$ and $p_\text{all swap}^\text{Pauli}$ are Pauli error probabilities combining all intermediate correction events and swappings, respectively. Using the identity
\begin{equation}
\sum_{\substack{j = 0\\j \text{ odd}}}^{k}\begin{pmatrix}k\\j\end{pmatrix}p^j(1-p)^{k-j}
 = \frac12\left[1 - (1-2p)^{k}\right]
\end{equation}  
for the sum of odd terms in a binomial distribution $\mathcal{B}(k, p)$, we can express the combined probabilities via the elementary probabilities:
\begin{equation}
\label{eq::allswap}
p_\text{all swap}^\text{Pauli} = \frac12 \left[1 - (1 - 2p_\text{swap}^\text{Pauli})^{(n-1)}\right]
\end{equation}
for the swapping, and
\begin{equation}
p_\text{all corr}^\text{Pauli} = \mathds{E}\left(\frac12 \left[1 - (1 - 2p_\text{corr}^\text{Pauli})^{\mathcal{M}_n}\right]\right)
\end{equation}
for the intermediate corrections. The expectation value can be evaluated using the relation in Eq.~(\ref{eq::expectation_approximation}) in the main text, which is derived in the last section of this Supplemental Material.
We then obtain
\begin{eqnarray}
 p_\text{all corr}^\text{Pauli} &=& \frac12\left[1 - (1 - 2p_\text{corr}^\text{Pauli})^{2m(n-1)}\left(\frac{1 - q}{1 + q}\right)^{n-1}\right.\nonumber\\
 &&\left.\times\left(\frac{1 + q(1 - 2p_\text{corr}^\text{Pauli})^m}{1 - q(1 - 2p_\text{corr}^\text{Pauli})^m}\right)^{n-1}\right].
\end{eqnarray}
Therein, $q$ denotes the entanglement distribution failure probability in each segment.

\subsection{Pauli errors from state generation}
As described in a later section of this Supplemental Material, we propose to generate Steane-GKP Bell pairs as well as hybrid entangled states containing a Steane-GKP component by connecting what we refer to as ``cube states'' with each other or with GKP single-photon hybrids via bare GKP BSMs. This process may lead to the introduction of Pauli errors due to finite-squeezing variance of the GKP modes, with the result that resource states for the repeater protocol, e.g. the logical Bell pair used for the loop-exit teleportation, are no longer perfect but contribute Pauli errors to the final QBER. Here we describe how to include such errors into the Steane-GKP scheme's rate analysis. 
The first step is to define the Pauli error probability $p^\text{Pauli}_\text{stategen}$ for each connecting BSM in the state generation via Eq.~(\ref{eq::paulierrorintegral}) with the variance chosen as $\sigma^2_\text{tot} = 2\delta^2$.
Since there is one possible error from each Steane-GKP Bell pair and each Steane-GKP hybrid state used for the protocol, the total number is given by the random variable
\begin{equation}
    \tilde{\mathcal{M}}_n = mD_n + 2(m+1)(n-1) = \mathcal{M}_n + 2(n-1),
\end{equation}
and we can modify Eq.~(\ref{eq::qber}) from the main text as
\begin{equation}
\label{eq::qber_steane}
    \epsilon = \left[\mathds{E}\left(\left(\mathbf{p}^\text{Pauli}_\text{stategen}\right)^{\circledast \tilde{\mathcal{M}}_n} \circledast
    \left(\mathbf{p}^\text{Pauli}_\text{corr}\right)^{\circledast \mathcal{M}_n}\right) \circledast \left( \mathbf{p}^\text{Pauli}_\text{swap}\right)^{\circledast n-1}\right]_1,
\end{equation}
where the vectors $\mathbf{p}^\text{Pauli}$ again stand for $[1 - p^\text{Pauli}, p^\text{Pauli}]$ for any given subscript, and the error probabilities from swapping and correction, $p^\text{Pauli}_\text{swap}$ and $p^\text{Pauli}_\text{corr}$, are given by passing the corresponding bare-GKP quantities defined by Eq.~(\ref{eq::paulierrorintegral}) through the Steane transfer function
\begin{equation}
q_\text{Steane}^\text{Pauli} = (q_\text{GKP}^\text{Pauli})^7 + 7(q_\text{GKP}^\text{Pauli})^6(1 - q_\text{GKP}^\text{Pauli}).
\end{equation}
As before, we can somewhat unravel this compact expression since we are working with qubits only. We begin by separating off the swapping contributions similarly to Eq.~(\ref{eq::qber_explicit}), since they contain no random variables:
\begin{equation}
    \epsilon = p_\text{all other}^\text{Pauli}(1 - p_\text{all swap}^\text{Pauli}) + p_\text{all swap}^\text{Pauli}(1 - p_\text{all other}^\text{Pauli}),
\end{equation}
with $p^\text{Pauli}_\text{all swap}$ given as in Eq.~(\ref{eq::allswap}) and $p^\text{Pauli}_\text{all other}$, collecting all events from state generation and intermediate corrections, given by the expectation value
 
\begin{eqnarray}
     &\mathds{E}&\left[\frac{1}{4}\left(1 - (1 - 2p^\text{Pauli}_\text{corr})^{\mathcal{M}_n}\right)\left(1 + (1 - 2p^\text{Pauli}_\text{stategen})^{\tilde{\mathcal{M}}_n}\right)\right.\nonumber\\
    &&+ \left.\frac{1}{4}\left(1 + (1 - 2p^\text{Pauli}_\text{corr})^{\mathcal{M}_n}\right)\left(1 - (1 - 2p^\text{Pauli}_\text{stategen})^{\tilde{\mathcal{M}}_n}\right) \right], \nonumber
\end{eqnarray}
which can be simplified to
\begin{equation}
    p^\text{Pauli}_\text{all other} = \mathds{E}\left[\frac{1}{2}\left(1 - (1 - 2p^\text{Pauli}_\text{corr})^{\mathcal{M}_n}(1 - 2p^\text{Pauli}_\text{stategen})^{\tilde{\mathcal{M}}_n}\right)\right].
\end{equation}
Using again our approximation formula from Eq.~(\ref{eq::expectation_approximation}) in the main text, we can evaluate the expectation to find
\begin{eqnarray}
    p^\text{Pauli}_\text{all other} &&= \frac12\left[\vphantom{\frac12^n}1 - (1 - 2p_\text{corr}^\text{Pauli})^{2m(n-1)}\right.\\
    &&\left.\times(1 - 2p_\text{stategen}^\text{Pauli})^{2(m+1)(n-1)}\right.\nonumber\\
    &&\times\left(\frac{1 - q}{1 + q}\right)^{n-1}\nonumber\\
    &&\left.\times\left(\frac{1 + q(1 - 2p_\text{corr}^\text{Pauli})^m(1 - 2p_\text{stategen}^\text{Pauli})^m}{1 - q(1 - 2p_\text{corr}^\text{Pauli})^m(1 - 2p_\text{stategen}^\text{Pauli})^m}\right)^{n-1}\right]\nonumber.
\end{eqnarray}

\subsection{QPC teleportation}
The $X$-basis code states of the $(b, a)$-QPC \cite{qpc} with $b$ blocks of $a$ photons each are given by
\begin{eqnarray}
    \ket{\overline{+}} &=& \frac{1}{\sqrt{2^n}}\left(\ket{0}^{\otimes a} + \ket{1}^{\otimes a}\right)^{\otimes b}\nonumber\\
    \ket{\overline{-}} &=& \frac{1}{\sqrt{2^n}}\left(\ket{0}^{\otimes a} - \ket{1}^{\otimes a}\right)^{\otimes b}, 
\end{eqnarray}
and the computational basis states are obtained from the usual relations $\ket{\overline{0}} = \sqnorm (\ket{\overline{+}} + \ket{\overline{-}})$ and $\ket{\overline{1}} = \sqnorm (\ket{\overline{+}} - \ket{\overline{-}})$.
In contrast to the GKP code, deterministic BSMs are not possible with the QPC; however, the success probability can be brought arbitrarily close to $1$ by increasing the number of blocks $b$.
When the physical qubits are realized as polarization-encoded dual-rail photons, i.e. $\ket{0} = \ket{H}$ and $\ket{1} = \ket{V}$, teleportation can be implemented transversally using only linear optics and photodetectors that can differentiate between zero, one and more than one photons. 
Each mode of the first logical state is combined with the corresponding mode of the other state at a regular beam splitter, whereupon the outgoing modes are split into horizontal and vertical components via polarizing beam splitters and subsequently measured by photodetectors placed at each of the four outputs. A Bell state is successfully detected if clicks are registered at two different detectors; otherwise, the BSM has failed. It is well-known \cite{Calsamiglia2001} that the probability of success of these physical BSMs is limited by $1/2$ without the use of auxillary photons, because only the states $\ket{\psi^+}$ and $\ket{\psi^-}$, or $\ket{\phi_{1, 0}}$ and $\ket{\phi_{1, 1}}$ in the notation we will adopt in this section, will result in distinguishable detector patterns. However, the structure of the QPC allows for a much higher success probability for BSMs on the logical level, even in the presence of photon loss. We will explain this in the following, summarizing the discussion in Refs.~\cite{Ewert2016, Ewert2017}. 

First, note that the QPC Bell states can, up to a reordering of modes (see Ref.~\cite{Ewert2016} for a proof and more detailed explanation), be written as 
\begin{eqnarray}
\label{eq::qpcbell}
    \ket{\overline{\phi_{k, l}}} &=& \sqnorm\left(\ket{\overline{0}}\ket{\overline{k}} + (-1)^l\ket{\overline{1}}\ket{\overline{1-k}}\right)\nonumber\\
    &=& \frac{1}{\sqrt{2^{b-1}}}\sum_{\mathbf{s}\in A_{k, b}}\bigotimes_{i=1}^{b}\ket{\phi_{s_i, l}^{(a)}},
\end{eqnarray}
where $A_{k, b} = \{\mathbf{s}\in\{0,1\}^b | \sum_{i=1}^b s_i = k\mod 2\}$ and the block-level Bell states $\ket{\phi_{k, l}^{(a)}}$ can be expressed as the superposition
\begin{equation}
\label{eq::blockbell}
    \ket{\phi_{k, l}^{(a)}} = \frac{1}{\sqrt{2^{a-1}}}\sum_{\mathbf{r}\in A_{l, a}}\bigotimes_{i=1}^{a}\ket{\phi_{k, r_i}},
\end{equation}
of products of the physical-level Bell states $\ket{\phi_{k, r_i}}$.
From Eq.~(\ref{eq::qpcbell}), it is clear that a logical Bell state can be correctly identified by finding $l$ and all $s_i$ if one succeeds in determining the second index in at least one block and the first index in each block. From Eq.~(\ref{eq::blockbell}), we recognize that finding the first block index requires only one physical BSM result per block, whereas finding the second index is only possible if all physical BSMs succeed (in stabilizer language \cite{gottesman_phd}, this corresponds to measuring the complete $X$ information in at least one block and some of the $Z$ information in each block \cite{Schmidt2019}). 
A physical BSM can only succeed if no photon is lost, such that we get the necessary condition of keeping at least one photon in each block and all photons in at least one block for success of the logical BSM. 
In the no-loss case, the first block index can always be identified, since the detectors will either produce a pattern that is unique to one of the states $\ket{\phi_{1, 0}}$ or $\ket{\phi_{1, 1}}$, or it will fail to do so, in which case we known that the first index is $0$. The second index can be found with probability $1/2$, namely always then when the first index is $1$.
The logical BSM success probability without loss thus reads
\begin{equation}
    p_\text{QPC}^\text{no loss} = 1 - \left(\frac{1}{2}\right)^b.
\end{equation}
To find the probability in the case of $\mu$ photon losses, we need to determine the number of possible ways to distribute the losses onto the $ab$ photons that are compatible with the condition of keeping at least one photon in each block and all photons in one block. This is done by introducing a variable $i$ which counts the number of blocks that are affected by loss, and then counting the number of possibilities of distributing these $i$ blocks among the total number of blocks, as well as of distributing the $\mu$ losses among the $i$ blocks. 
The first index will automatically be correctly identified if the aforementioned condition is fulfilled; however, finding the second index can still fail. Since only the $b-i$ unaffected blocks can yield information about the second index, one takes this into account by factoring in the no-loss success probability for a code with the reduced number of blocks $b-i$. 
Putting all of this together, one obtains 
\begin{equation}
\label{eq::pmuqpc}
    p_\text{QPC}^\mu = \frac{1}{\begin{pmatrix}ab\\\mu\end{pmatrix}}\sum_{i=0}^{\min(\mu, b-1)}\left(1 - 2^{-(b-i)}\right)\begin{pmatrix}b\\i\end{pmatrix}\sum_{\mathbf{j}\in B_{\mu, a, i}}\prod_{k=1}^i\begin{pmatrix}a\\j_k\end{pmatrix}
\end{equation}
with
$B_{\mu, a, i} = \{\mathbf{j}\in\{1, ..., a-1\}^i |\sum_{k=1}^i j_k = \mu\}$
as the success probability with $\mu$ losses.
The total success probability is constructed by multiplying $p_\text{QPC}^\mu$ with the probability of exactly $\mu$ losses occurring, and summing over $\mu$:
\begin{equation}
    p_\text{QPC} = \sum_{\mu=0}^{(b-1)(a-1)}p_\text{QPC}^\mu \begin{pmatrix}ab\\\mu\end{pmatrix}\eta^{ab-\mu}(1-\eta)^\mu.
\end{equation}
After inserting Eq.~(\ref{eq::pmuqpc}), this can be simplified to Eq.~(\ref{eq::qpcprob}) in the main text. For the details of the simplification, we again refer to Ref.~\cite{Ewert2016}.

\subsection{Entangled state generation}
In this section, we discuss the preparation of the entangled resource states required for our repeater scheme. This includes both the encoded Bell states used for loop teleportation as well as the hybrid states between an encoded state and a dual-rail qubit used for entanglement distribution.  
\subsubsection{GKP - GKP}
It is well-known that GKP Bell states can be created by interfering two so-called GKP qunaught states at a beam splitter \cite{Walshe2020}.
The ideal qunaught state $\ket{\varnothing}$ is defined similarly to the GKP logical states as a superposition of position quadrature eigenstates, but with a different spacing of $\sqrt{2\pi}$:
\begin{equation}
\ket{\varnothing} = \sum_{k \in \mathds{Z}}\Ket{k\sqrt{2\pi}}_q.
\end{equation}
The beam splitter performs the transformation
\begin{eqnarray}
    \ket{\varnothing}\ket{\varnothing} &=& \sum_{kl \in\mathds{Z}}\ket{l\sqrt{2\pi}}\ket{k\sqrt{2\pi}} \nonumber\\
    &\rightarrow& \sum_{kl\in\mathds{Z}}\ket{(k+l)\sqrt{\pi}}\ket{(k-l)\sqrt{\pi}},
\end{eqnarray}
yielding the desired state, since $k+l$ and $k-l$ always have the same parity, and thus we have a superposition of both modes in $\ket{\overline{0}}$ or both in $\ket{\overline{1}}$.
For the creation of GKP states, whether qunaught or logical, there exist several ideas. In the original GKP paper \cite{gkp} it was proposed to have a momentum-squeezed vacuum state interact with a coherent state via a controlled phase rotation, followed by a homodyne measurement of the coherent state's momentum (see also Ref.~\cite{Budinger2024}). For optical systems, however, this is not easy since the required nonlinearity is hard to implement in practice. Other proposals for GKP creation avoiding the need for nonlinear interactions have followed, such as cat breeding \cite{catbreeding, Weigand2018}, boson sampling \cite{gkp_bosonsampling} and Gaussian breeding \cite{gkp_gaussianbreeding}.

\subsubsection{GKP - Single Photon}
\begin{figure}
\includegraphics[width=0.3\textwidth]{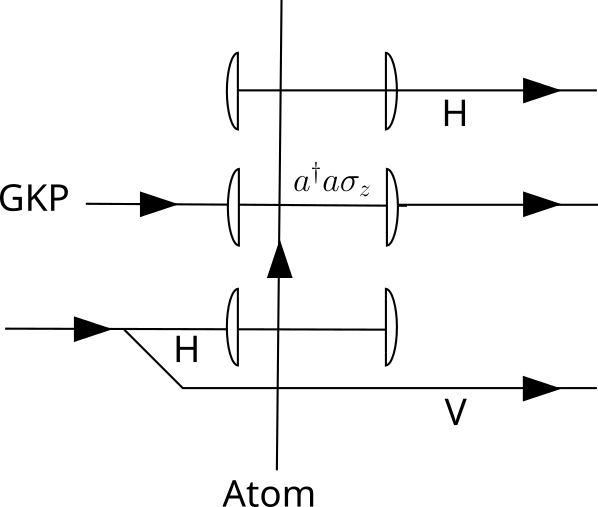}
\caption{Entangling a single-photon dual-rail qubit with a GKP qubit using cavity QED. }
\label{fig::hybridstate_cqed}
\end{figure}

For the GKP hybrid Bell state we present two possible generation schemes, one based on cavity QED and the other as part of a cat breeding protocol.
We begin with the former, which is illustrated in Fig.~\ref{fig::hybridstate_cqed}.
As input states we consider a dual-rail single-photon qubit starting in the state $(\ket{H} + \ket{V})/\sqrt{2}$ and a GKP qubit initialized in $\ket{\overline{0}}$. First, the single photon passes through a polarizing beam splitter that separates the polarization components onto different propagation directions. While the vertical component bypasses the cavity, the horizontal component interacts resonantly via $a_H\sigma_+ + a_H^\dagger\sigma_-$ with an atom modeled as a two-level system with states $\ket{g}$, $\ket{e}$ and jump operators $\sigma_+ = \ket{e}\bra{g}$, $\sigma_- = \ket{g}\bra{e}$, and starting in the ground state $\ket{g}$. The result of this interaction is the combined state
\begin{equation}
(\ket{0}\ket{e} + \ket{V}\ket{g})/\sqrt{2}, 
\end{equation} 
where the horizontal component has been absorbed by the atom.
Afterwards, the atom enters the second cavity, where it interacts dispersively with the GKP state. The interaction $a^\dagger a \sigma_z$ implements a controlled phase rotation; however, we actually want to shift conditioned on the atom being in the ground state. This can be achieved by performing a shift by $\sqrt{\pi}/2$ in positive $q$-direction and a $\pi/2$-rotation on the GKP state prior to the interaction with the atom, and afterwards performing another shift by $\sqrt{\pi}/2$.
After this, the combined state of all three systems reads
\begin{equation}
(\ket{0}\ket{e}\ket{\overline{0}} + \ket{V}\ket{g}\ket{\overline{1}})/\sqrt{2}.
\end{equation}
The atom now passes through the third cavity, where it again interacts resonantly with the horizontal mode, transferring the excitation back to the electromagnetic field, and producing the desired final state
\begin{equation}
\sqnorm\left(\ket{H}\ket{\overline{0}} + \ket{V}\ket{\overline{1}}\right)\ket{g},
\end{equation}
where we factored out the atom's state to the end.
While this method works in theory, it is not particularly well suited for experimental implementation due to issues with stabilization of the cavities or spontaneous emission destroying the atomic excitation.  \\

A more practical approach is our second method, where we start with a partially bred GKP state with a $q$-quadrature spacing of $2\sqrt{2\pi}$, i.e. $\sum_k\ket{2\sqrt{2\pi}k}_q$, and a hybrid entangled state between a single photon and squeezed coherent states $(\ket{H}\ket{a/2}_q + \ket{V}\ket{-a/2}_q)/\sqrt{2}$, which can be obtained probabilistically with linear optics according to Ref.~\cite{Kiryu2025}. For simplicity, here we have replaced the squeezed coherent states with $q$-eigenstates. First, the continuous-variable component of the hybrid state is displaced to change it into $(\ket{H}\ket{0}_q + \ket{V}\ket{-a}_q)/\sqrt{2}$, bringing the total initial state to
\begin{equation}
\frac{1}{\sqrt{2}}(\ket{H}\ket{0}_q + \ket{V}\ket{-a}_q)\sum_k\ket{2\sqrt{2\pi}k}_q.
\end{equation}
A beam splitter between the GKP and the squeezed coherent mode changes it to
\begin{eqnarray}
&&\frac{1}{\sqrt{2}}\left(\ket{H}\sum_k\ket{2\sqrt{\pi}k}_q\ket{2\sqrt{\pi}k}_q\right.\nonumber\\
&+& \left.\ket{V}\sum_k\ket{2\sqrt{\pi}k - a/\sqrt{2}}_q\ket{2\sqrt{\pi}k + a/\sqrt{2}}_q\right).
\end{eqnarray}
Measuring the $p$-quadrature of the second mode and postselecting to a multiple of $2\sqrt{\pi}$ in order to avoid phase factors results in
\begin{equation}
\frac{1}{\sqrt{2}}\left(\ket{H}\sum_k\ket{2\sqrt{\pi}k}_q + \ket{V}\sum_k\ket{2\sqrt{\pi}k + a/\sqrt{2}}_q\right)
\end{equation}
and together with the choice $a = \sqrt{2\pi}$ this yields the desired state 
\begin{eqnarray}
&&\frac{1}{\sqrt{2}}\left(\ket{H}\sum_k\ket{2\sqrt{\pi}k}_q + \ket{V}\sum_k\ket{(2k+1)\sqrt{\pi}}_q\right) \nonumber\\
&=& \frac{1}{\sqrt{2}}(\ket{H}\ket{\overline{0}} + \ket{V}\ket{\overline{1}}).
\end{eqnarray}

The postselection can be avoided if feedforward and a method of introducing variable relative phases between the polarization components are available:
The post-measurement state with an arbitrary measurement result $\tilde{p}$ is
\begin{eqnarray}
&&\frac{1}{\sqrt{2}}\left(\ket{H}\sum_k e^{-i2\sqrt{\pi}k\tilde{p}}\ket{2\sqrt{\pi}k}_q\right. \nonumber\\
&+&\left. \ket{V}\sum_k e^{-i(2\sqrt{\pi}k - a/\sqrt{2})\tilde{p}}\ket{2\sqrt{\pi}k + a/\sqrt{2}}_q\right),
\end{eqnarray}
which can be brought to
\begin{equation}
    \frac{1}{\sqrt{2}}\left(\ket{H}\sum_k \ket{2\sqrt{\pi}k}_q + \ket{V}\sum_k e^{ia\sqrt{2}\tilde{p}}\ket{2\sqrt{\pi}k + a/\sqrt{2}}_q\right)
\end{equation}
by application of a momentum shift $e^{i\tilde{p}q}$.
With the choice $a = \sqrt{2\pi}$ this becomes $(\ket{H}\ket{\overline{0}} + e^{i2\sqrt{\pi}\tilde{p}}\ket{V}\ket{\overline{1}})/\sqrt{2}$, which is almost the desired state, but with an unwanted relative phase that depends on the measurement result $\tilde{p}$. It is difficult to cancel this phase on the GKP modes; however, it might be possible to do so on the single-photon components using feedforward and a device that can introduce tunable phase differences between horizontal and vertical polarizations, e.g. a Pockels cell.

\subsubsection{Steane-GKP - Steane-GKP}
The key insight for the preparation of Steane-GKP Bell states is the fact that an eight-qubit cluster state in the shape of a cube is, up to local Hadamard operations, equivalent to a Bell state formed by a logical Steane-qubit (comprised of seven ``physical'' qubits) and a bare qubit \cite{Rozpedek2023, Cafaro2014}.
In our case, the individual qubits are realized as GKP qubits, such that effectively we have the state
\begin{equation}
    \sqnorm\left(\ket{\overline{\overline{0}}}\ket{\overline{0}} + \ket{\overline{\overline{1}}}\ket{\overline{1}}\right),
\end{equation}
where kets with a double overline denote code states of the concatenated Steane-GKP code and kets with a single overline denote GKP qubits. 
Taking two copies of such a state, we can swap the entanglement by performing a BSM on the bare GKP modes, leaving the Steane-GKP components in a Bell state.
The cube state could in principle be created by initializing eight GKP modes in the state $\ket{\overline{+}}$ and connecting them with logical CZ gates $e^{-i\sqrt{\pi}q_1q_2}$ along the cube's edges. However, this is not ideal since CZ gates lead to an unwanted accumulation of variance in the finite-squeezing case, negatively impacting the final state's quality. Instead, Ref.~\cite{Rozpedek2023} envisions a scheme where the role of CZ gates is limited to the production of GHZ states that are then successively stitched together to form the cube via fusion measurements. These can be represented as a composition of a Hadamard gate and a regular BSM, and can therefore be implemented with linear optics for GKP qubits. 
Another, more efficient option is proposed in Ref.~\cite{cubecreation}. It allows the creation of cube states using only passive linear optics and a minimal resource overhead.

\subsubsection{Steane-GKP - Single Photon}
The hybrid state of a Steane GKP qubit and a dual-rail qubit needed for entanglement distribution in the Steane-GKP repeater can be created similarly via entanglement swapping between a cube state and a bare-GKP hybrid state by combining one mode of the cube with the GKP mode of the hybrid state at a beam splitter and homodyning the outputs.

\subsubsection{QPC - QPC}
In Ref.~\cite{Ewert2017}, two schemes for generating QPC Bell states were presented: One deterministic using non-linear interactions, and one probabilistic with only linear interactions. The former is based on coherent photon conversion \cite{Langford2011}, a technique that allows for the transformation
\begin{equation}
    \alpha\ket{H} + \beta\ket{V} \rightarrow \alpha\ket{H}\ket{H} + \beta\ket{V}\ket{V},
\end{equation}
referred to as photon doppling. Together with half-wave plates realizing Hadamard gates on polarization-encoded qubits, transformations of this form can be used to turn $\ket{H}$ into a QPC-$\ket{\overline{0}}$ state and $\ket{V}$ into a QPC-$\ket{\overline{1}}$ state, and thus to implement the chain of transformations
\begin{eqnarray}
    \sqnorm(\ket{H} + \ket{V}) &\rightarrow& \sqnorm(\ket{H}\ket{H} + \ket{V}\ket{V})\nonumber\\
    &\rightarrow& \sqnorm (\ket{\overline{0}}\ket{\overline{0}} + \ket{\overline{1}}\ket{\overline{1}}).
\end{eqnarray}\\

The other scheme makes use of repeated BSMs to generate the state
\begin{equation}
    \Ket{B^{(b, a)}} = \sqnorm\left(\ket{\overline{0}}\ket{\phi_{0, 0}} + \ket{\overline{1}}\ket{\phi_{0, 1}}\right)
\end{equation}
from GHZ states $\sqnorm(\ket{H}\ket{H}\ket{H} + \ket{V}\ket{V}\ket{V})$, that can in turn be probabilistically created by linear optics.
A measurement of the last photon in the $X$-basis, possibly followed by a Pauli correction depending on the result, yields the state
\begin{equation}
\label{eq::Bx}
    \Ket{B^{(b, a)}_X} = \sqnorm\left(\ket{\overline{0}}\ket{+} + \ket{\overline{1}}\ket{-}\right),
\end{equation}
and the desired encoded Bell state can be obtained from two copies of such a state by a BSM on the respective last photons.

\subsubsection{QPC - Single Photon}
The second method of generating QPC Bell pairs also enables production of the hybrid entangled state $(\ket{H}\ket{\overline{0}} + \ket{V}\ket{\overline{1}})/\sqrt{2}$. In fact, all that is needed to transform the state $\ket{B^{(b, a)}_X}$ in Eq.~(\ref{eq::Bx}) to the target state is a Hadamard gate on the last photon and a reordering of modes.

\subsection{Operational regimes: short vs. long segments}
\begin{figure*}
\begin{minipage}{\textwidth}
\includegraphics[width=0.95\linewidth]{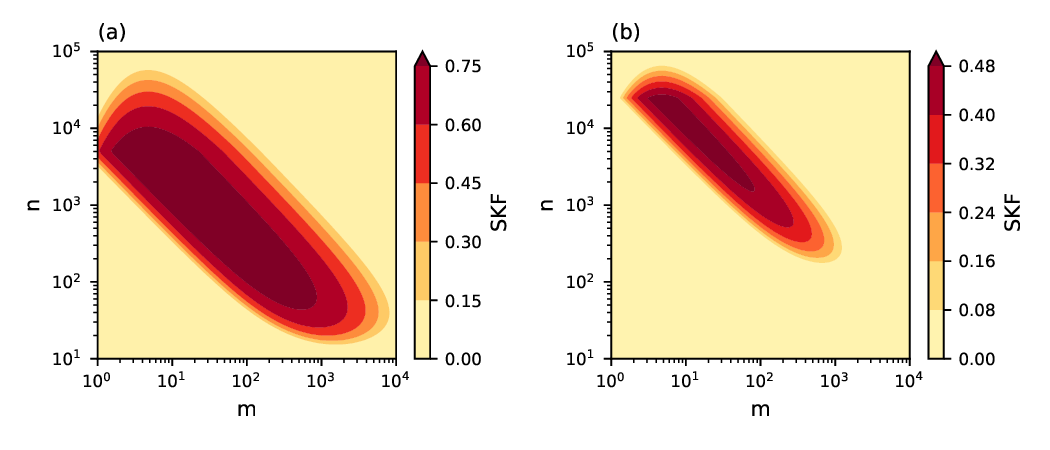}
\end{minipage}

\begin{minipage}{\textwidth}
\includegraphics[width=0.95\linewidth]{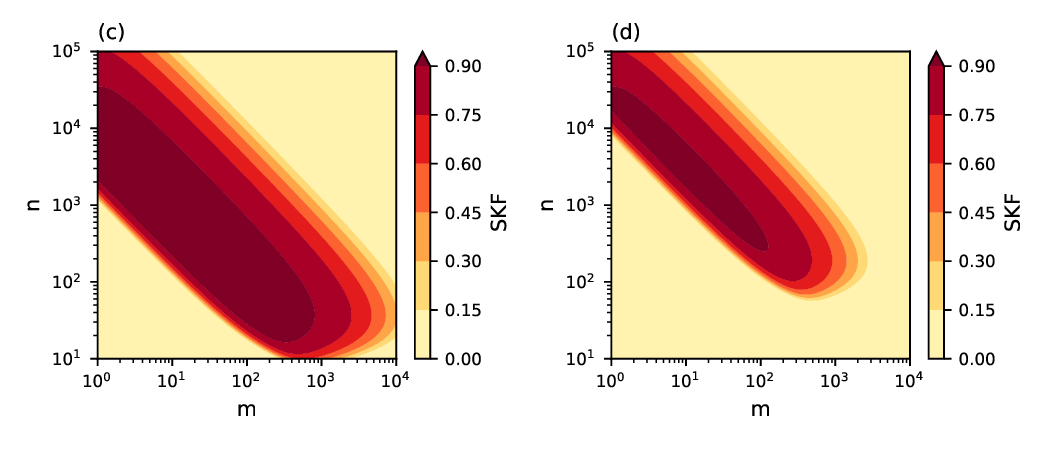}
\end{minipage}

\begin{minipage}{\textwidth}
\includegraphics[width=0.95\linewidth]{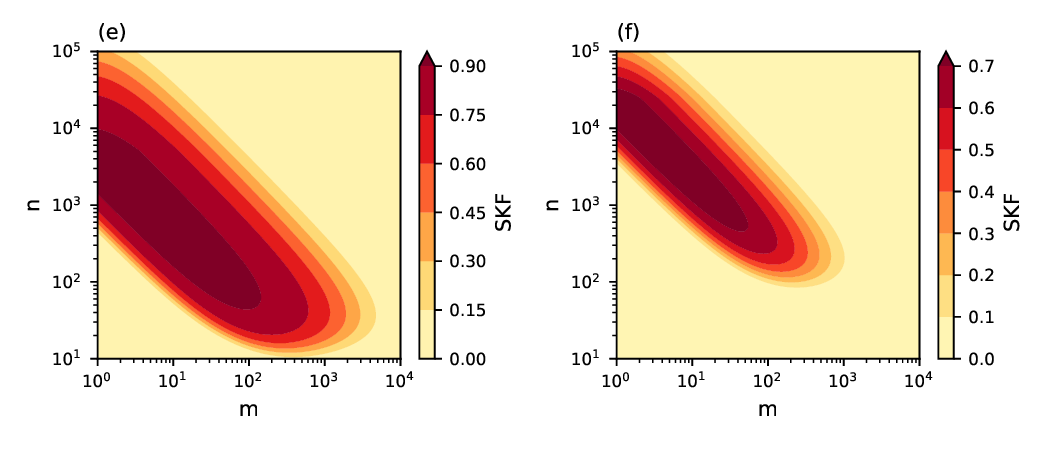}
\end{minipage}
\caption{(Color online) Secret key fraction plotted in the $n$-$m$-plane for the three repeater schemes. Plots for Steane-GKP include the effect of Pauli errors during state preparation. (a) GKP (18dB) $L=1000$km, (b) GKP (18dB) $L=5000$km, (c) Steane-GKP (16dB) $L=1000$km, (d) Steane-GKP (16dB) $L=5000$km, (e) QPC (21,5) $L=1000$km, (f) QPC (21,5) $L=5000$km. All plots use $\plink=\ploop=0.99$.}
\label{fig::mn_contours}
\end{figure*}

When plotting the SKF over the $n$-$m$-plane for the different schemes, one observes that the areas where a non-vanishing SKF can be achieved take the shape of elongated ellipses oriented roughly along the diagonal connecting the upper left with the lower right corners of the plots, as illustrated in Fig.~\ref{fig::mn_contours}. Therein, the first, second and third row corresponds to GKP, Steane-GKP and QPC repeaters, and the left and right column to a total distance of $L=1000$km and $L=5000$km, respectively.
This demonstrates that both a high $n$, by increasing the entanglement distribution success probability, and a high $m$, by reducing the accumulation of noise during wait times, are suitable to address memory loss; however, choosing both parameters high results in the presence of too many sources of Pauli errors, such that even small noise at each individual event overwhelms the SKF.
Considering the ellipses' extremal points gives rise to two complementary operational regimes: one in the upper left corner, characterized by a low $m$ but a relatively high number of segments, and another near the lower right corner with fewer segments and more intermediate correction steps. In the former, the repeater consumes much fewer resource states and offers an advantage in the form of a better raw rate; however, the required number of stations is several orders of magnitude higher than in the latter. In a practical application, one would therefore have to consider whether state generation or additional intermediate stations are more costly, and choose the regime accordingly.

\subsection{Absolute value of geometric random variables' difference}
Since the entanglement distribution process in each segment of the repeater follows a geometric distribution, the number of timesteps $W$ that modes of neighboring segments have to wait for each other corresponds to the absolute value of the difference of two geometric random variables. 
Let $\mathcal{N}_1$ and $\mathcal{N}_2$ be two geometrically distributed random variables with success probability $p$. In this section, we will derive the probability distribution as well as the expectation value of  $|\mathcal{N}_1 - \mathcal{N}_2|$. \\

The difference $\mathcal{N}_1 - \mathcal{N}_2$ takes the value $k\in\mathds{Z}$ whenever $\mathcal{N}_1 = n$ and $\mathcal{N}_2 = n-k$ for any $n$ compatible with the condition $\mathcal{N}_{1, 2} \geq 1$, which must be fulfilled for $\mathcal{N}_1$ and $\mathcal{N}_2$ to be valid geometric random variables.
This condition is fulfilled whenever $n \geq \max(1, k+1)$, and thus the probability is given by
\begin{eqnarray}
\mathds{P}(\mathcal{N}_1 - \mathcal{N}_2 = k) &=  &\sum_{n = \max(1, k+1)}^\infty \mathds{P}(\mathcal{N}_1 = n)\mathds{P}(\mathcal{N}_2 = n-k) \nonumber\\
&= &\sum_{n = \max(1, k+1)}^\infty pq^{n-1}pq^{n-k-1}\nonumber\\
& = &\frac{p^2}{q^{2+k}}\sum_{n = \max(1, k+1)}^\infty (q^2)^n,
\label{app::eq::stat1}
\end{eqnarray}
where $q = 1-p$. 
The sum in Eq.~(\ref{app::eq::stat1}) is reminiscent of a geometric series; however, the lower bound is not 0 and the sum needs to be rewritten as $\sum_{n=m}^\infty = \sum_{n=0}^\infty - \sum_{n=0}^{m-1}$ before the well-known results about the limit and the finite partial sums of the geometric series can be applied. At this stage, it makes sense to introduce a case distinction between positive and strictly negative~$k$,
\begin{equation}
    \mathds{P}(\mathcal{N}_1 - \mathcal{N}_2 = k) = \frac{p^2}{q^{2+k}}
    \begin{cases}
        \sum_{n = k+1}^\infty (q^2)^n & k\geq 0\\
        \sum_{n = 1}^\infty (q^2)^n & k < 0
    \end{cases}
\end{equation}
and then perform the rewrite on both cases:
\begin{equation}
\mathds{P}(\mathcal{N}_1 - \mathcal{N}_2 = k) =  \frac{p^2}{q^{2+k}} 
\begin{cases}
\left[\frac{1}{1-q^2} - \frac{1-q^{2k+2}}{1-q^2}\right] & k\geq 0 \\
\left[\frac{1}{1-q^2} - 1\right] & k < 0.
\end{cases}
\end{equation}
Simplifying, one arrives at
\begin{equation}
\mathds{P}(\mathcal{N}_1 - \mathcal{N}_2 = k) =  
\begin{cases}
\frac{p^2q^k}{1 - q^2} & k\geq 0 \\
\frac{p^2}{q^k(1 - q^2)} & k < 0,
\end{cases}
\end{equation}
from which the distribution of the difference's absolute value can be read off immediately:
\begin{equation}
\label{app::eq::probdist}
\mathds{P}(|\mathcal{N}_1 - \mathcal{N}_2| = k) = \begin{cases}\frac{p^2}{1-q^2} & k = 0\\
\frac{2p^2q^k}{1-q^2} & k > 0.\end{cases}
\end{equation}

\subsection{Expectation value for exponential of summed waiting time}
In this section, we derive the relation in Eq.~(\ref{eq::expectation_approximation}) of the main text, which is an approximation for the expectation value of the exponential summed waiting time.
We proceed in two steps: First, consider the fact that $D_n$ can be written as a sum of $n-1$ random variables describing the number of time steps that the state belonging to the segment that finishes distribution first has to wait for the other segment at each of the repeater stations. In general, these random variables are not all mutually independent \cite{rateanalysis, Goodenough2025}; however, we will make the approximation that they can be treated as independent \cite{häussler_vanloock}. 
Each of them is then given by the absolute value of the difference between two geometric random variables with the probability distribution 
\begin{equation}
\label{eq::waitdist}
\mathds{P}(|\mathcal{N}_1 - \mathcal{N}_2| = k) = \begin{cases}\frac{p^2}{1-q^2} & k = 0\\
\frac{2p^2q^k}{1-q^2} & k > 0\end{cases}
\end{equation}
derived in the preceding section,
where $p$ and $q=1-p$ are the entanglement distribution and failure probabilities in one segment, respectively. 
In the following, we will denote such a random variable with $W$.
Under the assumption of independence, we can now write
\begin{equation}
    \mathds{E}\left(a^{D_n}\right) = \mathds{E}\left(a^{\sum_{i=1}^{n-1}W}\right) = \mathds{E}\left(\prod_{i=1}^{n-1}a^W\right) = \left[\mathds{E}\left(a^W\right)\right]^{n-1},
\end{equation}
such that, as the second step, it suffices to calculate $\mathds{E}(a^W)$ using the aforementioned probability distribution:
\begin{eqnarray}
    \mathds{E}(a^W) &=& \sum_{k=0}^\infty a^k\mathds{P}(W=k) = \frac{p^2}{1 - q^2} + \sum_{k=1}^\infty a^kq^k\frac{2p^2}{1 - q^2}\nonumber\\
    &=& \frac{p^2}{1-q^2}\left(1 + 2\sum_{k=0}^\infty (aq)^k - 2\right)\nonumber\\
    &=& \frac{p^2}{1-q^2}\left(\frac{2}{1 - aq} - 1\right),
\end{eqnarray}
where the last equality is valid if $|a| < 1$.
After some minor rewrites we obtain 
\begin{equation}
    \mathds{E}(a^W) = \frac{1-q}{1+q}\frac{1+aq}{1-aq}
\end{equation}
and therefore
\begin{equation}
    \mathds{E}\left(a^{D_n}\right) = \left(\frac{1-q}{1+q}\frac{1+aq}{1-aq}\right)^{n-1}
\end{equation}
as in Eq.~(\ref{eq::expectation_approximation}).

\end{document}